\documentclass[10pt,conference]{IEEEtran}
\IEEEoverridecommandlockouts

\usepackage{cite}
\usepackage{amsmath,amssymb,amsfonts}
\usepackage{algorithmic}
\usepackage{graphicx}
\usepackage{textcomp}
\usepackage{xcolor}
\def\BibTeX{{\rm B\kern-.05em{\sc i\kern-.025em b}\kern-.08em
    T\kern-.1667em\lower.7ex\hbox{E}\kern-.125emX}}

\usepackage{pbalance}

\makeatletter
\newcommand{\linebreakand}{%
  \end{@IEEEauthorhalign}
  \hfill\mbox{}\par
  \mbox{}\hfill\begin{@IEEEauthorhalign}
}
\makeatother
\begin{document}
\title{A Close Reading Approach to Gender Narrative Biases in AI-Generated Stories}

\author{Daniel Raffini, Agnese Macori, Marco Angelini, Tiziana Catarci\thanks{Daniel Raffini, Agnese Macori, Tiziana Catarci are with Sapienza University of Rome, Italy: E-mail: \{raffini, macori, catarci\}@diag.uniroma1.it. Marco Angelini is with Link University of Rome, Italy: E-mail: m.angelini@unilink.it}}

\maketitle

\begin{abstract}
The paper explores the study of gender-based narrative biases in stories generated by ChatGPT, Gemini, and Claude. The prompt design draws on Propp’s character classifications and Freytag’s narrative structure. The stories are analyzed through a close reading approach, with particular attention to adherence to the prompt, gender distribution of characters, physical and psychological descriptions, actions, and finally, plot development and character relationships. The results reveal the persistence of biases — especially implicit ones — in the generated stories and highlight the importance of assessing biases at multiple levels using an interpretative approach.

\end{abstract}


\vspace{-4mm}
\section{Introduction}
\label{sec:intro}

In recent years, considerable attention has been paid to addressing the problem of bias in Large Language Models (LLMs). Despite ongoing efforts and improvements, LLMs still often do not adequately represent diversity and continue to propagate various forms of societal bias in their output \cite{Gallegos} \cite{Torres} \cite{Zhou}. The extensive use of LLMs for content creation and text generation makes this issue increasingly urgent. Regarding gender bias, studies have explored different aspects, such as the correlation between gender and occupation \cite{Kotek} \cite{Doll}, personas \cite{Ranjan1} \cite{Cheng}, or the use of adjectives \cite{Sounda}. Many of these studies also compared LLMs' correlations with official social data on occupation and human perceptions \cite{Doll} \cite{Bas}. 

Methodologies for studying bias in LLMs can be divided into intrinsic and extrinsic approaches \cite{Zayed} \cite{Li}. The intrinsic approach includes embedding- and probability-based bias, while the extrinsic approach focuses on generation-based bias \cite{Chu}. A recent study from UNESCO \cite{Unesco} provides a comprehensive application of various approaches by studying the connection of gendered words, asking LLMs to complete sentences, and generating entire stories.

There are different modes of gender bias and stereotype propagation, and it is important to evaluate the issue from various points of view. Scholars have designed different bias taxonomies based on factors such as origin, subjects involved, and social and operational implications \cite{Navigli} \cite{Ranjan2}. From the standpoint of communication and narrative studies, we propose distinguishing three types:
\begin{enumerate}
\item \textbf{Linguistic bias}, which arises from the use of certain language characteristics, such as the correlation of extended masculine or gender-coded words. This type of bias can be studied, for example, by analyzing word embeddings or co-occurrences.
\item \textbf{Interpretative bias}, when bias affects the understanding of a text and influences its interpretation. This applies to tasks such as summarizing, text analysis, information extraction, classification, and answering statements-based questions.
\item \textbf{Narrative bias}, when stereotypes emerge not from a single linguistic element, but from a narrative that involves multiple passages, descriptions, and actions. This bias typically arises through the free generation of stories in response to a specific prompt.
\end{enumerate}

In our study, we focus on narrative bias. Among existing methodologies, open-ended generation is the most suitable for analyzing narrative bias, as it allows us to examine bias development throughout entire texts, rather than just in short sentences or single words. In their study, UNESCO researchers asked models to generate stories about boys, girls, women, and men, and then created a word cloud for each category, revealing stereotypical differences in the setting of the story and the adjectives used. The results showed some interesting features; for example, in stories about women, husbands were mentioned more frequently than wives in stories about men, and women were associated with stereotypical roles and settings. The study also revealed that family stereotypes were prevalent when LLMs were asked to place the story in the global South, while love was the main theme associated with women in narratives set in the Global North, suggesting that gender stereotypes may vary between cultural identities. The analysis was carried out on 1000 samples for each category, and the stories were analyzed using a computational approach \cite{Unesco}.

Computational and distant reading approaches to the analysis of gender bias in LLM-generated texts are prevalent \cite{Kumar}. Most existing studies use computational techniques such as classifiers \cite{Sounda} or topic modeling \cite{Lucy}. In our study, we propose a close reading, focusing on a smaller sample but offering in-depth analysis of characters' representation from a gender perspective, using literary criticism, narratology theories, and rhetorical analysis combined with gender studies \cite{Smith} \cite{Brummett}. We believe a human approach is useful for uncovering implicit biases, which are not always evident at the linguistic level and require a connection to social and rhetorical context. This approach also addresses composite bias, which results from the combination of different parts of the narration.

In our study, we focus on narrative characters in LLM-generated stories. We will analyze their functions, descriptions, and actions in relation to gender. Finally, we will address plot analysis. \cite{b9} and \cite{Lucy} demonstrate the usefulness of character analysis in AI story generation, but they focus only on the main character. In our study, we include different types of characters, following Propp’s classification \cite{b8}, to investigate narrative bias in the distribution, connotation, and relationships among different types of characters in the text. Addressing gender bias in AI-generated stories is essential, given the importance of storytelling in our communication and the growing use of LLMs to generate textual content and narratives, which can impact and shape our imagination and society. The perpetuation of bias can affect the realm of narrative and imagination, reinforcing stereotypes that societies are working to change. Focusing on narrative bias is particularly interesting because implicit stereotypes conveyed through narration are more likely to influence human beliefs \cite{Caliskan}. Our study aims to evaluate the presence of gender bias in stories generated by three widely used LLMs: ChatGPT, Gemini, and Claude. At this stage, we focus on analysis, while future research may involve comparing the results of this analysis with the most recent mitigation techniques.

The main contributions of our study are:
\begin{itemize}
    \item The investigation of gender bias in stories generated by LLMs through a comprehensive and close reading approach, expanding on previous studies that focused on specific aspects or primarily computational methods.
    \item A detailed human-centered analysis of narrative biases in the texts, highlighting the distinctive tendencies of the three models under examination in terms of character representation, characterization, and plot development.
\end{itemize}

The "Related Work" section contextualizes the study within the previous bibliography on gender representation in both traditional and AI-generated narratives. The "Methodology" outlines the experimental setup, including prompt design based on Propp’s character functions and Freytag’s plot structure, and the use of close reading for analysis. The "Results" section presents findings on five analytical levels: adherence to the prompt, gender distribution, character descriptions, character actions, and plot dynamics. The "Discussion" interprets these findings, comparing the biases and narrative strategies across ChatGPT, Gemini, and Claude. Finally, the "Conclusion" summarizes the insights, highlights methodological limitations, and proposes directions for future research.

\section{Related Work}
\label{sec:narrative}

Gender bias has implications for both narrative texts generated by LLMs and those produced by humans over the course of centuries. LLMs acquire knowledge of the world through their training data and replicate stereotypes and oversimplifications ~\cite{b10} ~\cite{b11}. Therefore, it is necessary to focus attention on studies that investigated the presence of bias within narratives. Among them, fairy tales prove useful, because they represent one of the most elementary and schematizable narrative forms \cite{b8}. In their analysis of Grimm’s fairy tales, Baker-Sperry and Grauerholz~\cite{b1} highlighted the prevalence of the ideal of feminine beauty and its correlation with the longevity of these narratives. This suggests a historical reinforcement of this ideal. Gottschall~\cite{b2} addressed the limited research on female protagonists by performing a quantitative analysis of global folktales to identify cross-cultural patterns in their characterization. Ragan's statistical analysis \cite{b3} of a substantial multicultural collection of folktales highlighted the impact of gender on the compilation and interpretation of these narratives. Gottschall~\cite{b4} cross-cultural study of folktales revealed a nearly-universal emphasis on female attractiveness. Francemone~\cite{b5} research on the formation of audience disposition posited that, while character schemes operate similarly across genders, moral evaluations of behavior elicit stronger affective responses towards female characters. Nusbaumer~\cite{b6} comparative analysis of \textit{Cinderella} and \textit{Love in a Fallen City}~\cite{b7} explored the concept of the "Cinderella complex", positing that both narratives, despite potential feminist interpretations, perpetuate female dependence on male intervention. To summarize, related literature reveals a consistent historical trend in human narratives, characterized by emphasis on female physical attractiveness and the portrayal of female characters within specific, often dependent, roles. The presence of established patterns in human-authored texts provides a significant framework for evaluating the potential for the replication or subversion of such gendered representations in contemporary narratives generated by LLMs.

The pervasiveness of gender-connoted roles is such that Propp formalized them in a classification of fairy tale characters \cite{b8}. Propp is a significant figure in literary structuralism, an approach that aims to identify the underlying structures and patterns in narrative texts by analyzing elements — such as characters, actions, and themes — to discern the rules and conventions governing narration. From this theoretical framework emerged narratology, a broader field of study concerned with theories and models of narrative. By generalizing from textual analysis, Propp inductively proposes the identification of typical characters within fairy tales in his work, \textit{Morphology of the Folktale}~\cite{b8}: the villain, the donor, the helper, the princess (and her father), the dispatcher, the hero, the false hero. From the perspective of gender stereotypes it is important to note that Propp's formalization classifies the protagonist as male and the object of desire as female. The princess is also associated with a male figure, the father, who is capable of protecting and safeguarding her until the hero's arrival and who will marry her. Propp’s methodology is inductive, essentially discerning common characteristics across a corpus of folktales. Consequently, it becomes evident that a significant proportion of these narratives portray women as vulnerable subjects and objects of male desire and gaze. 

Propp does not actually classify characters themselves, but rather the spheres of action that pertain to each role. He proposes a division of the fairy tale structure into 31 functions~\cite{b8} representing individual actions that, when combined, constitute the backbone of these narratives. Several more accessible proposals have adapted and condensed this schema, notably those of Greimas~\cite{b12} and Todorov~\cite{b13}. Greimas reduced the schema to six elements, and Todorov explained that the fundamental element of a narrative is the transition from one state of equilibrium to another~\cite{b14}. For the purposes of our study, the clearest and most effective schematization is Freytag’s pyramid~\cite{b15}, who provide a graphic representation of the five essential plot points. Drawing upon the partition proposed by Aristotle in \textit{The Poetics}~\cite{b16}, this model subdivides the plot into an initial \textit{introduction}, a series of complications and transforming actions that negatively shift the situation (\textit{the rise}), a point of suspense representing the \textit{climax}, a phase of resolving the initial difficulties (\textit{the fall}), and a denouement in which the hero’s destiny is fulfilled (\textit{the catastrophe}). Although this schema is primarily associated with drama, it is useful for schematizing the typical progression of traditional narratives~\cite{b15}.

\begin{figure}
    \centering
    \includegraphics[width=0.50\linewidth]{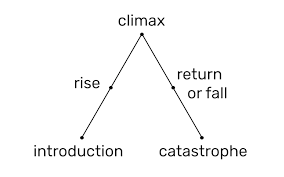}
    \caption{Freytag's pyramid}
    \label{fig:enter-label}
\end{figure}

In contrast to Propp's granular framework, Freytag's model does not impose rigid constraints on the relationship between characters and their actions. This distinction is essential when transitioning from the hermeneutic phase, which involves understanding existing narratives, to the creative stage of producing original texts. Propp's model, particularly when used as an input for generative AI, has the potential to restrict characters' actions based on their designated function. Conversely, Freytag's pyramid establishes specific conditions for plot development while allowing for greater variability in character creation and agency. This flexibility is essential for promoting originality and preventing the replication of formulaic character behaviors that could result from Propp's restrictive model. 

\section{Methodology} 
\label{sec:methodology}

Among the available methodologies for analyzing gender bias in LLMs, we chose open-ended story generation through a structured prompt to address narrative bias in relation to characters \cite{Chen}. In this context, careful attention to the prompt definition is necessary, as previous studies have demonstrated \cite{Doll}. We opt for zero-shot prompting \cite{Sounda}, aiming to allow text generation to be free in order to better understand the production of gender bias, which could be influenced by examples. The prompt was designed to provide instructions on the functions of the characters and the text structure, drawing on relevant and well-known narrative models, namely Propp's functions for characters and Freytag’s story structure. Due to the nature of short stories, Propp’s original seven character types have been reduced and adapted. We chose to rename the hero as the main character and the princess as the desired character, aiming for more neutral terminology. Additionally, we combined the helper and the donor into a single role and eliminated the false hero, as including this figure would have introduced excessive narrative complexity given the limited length of the texts. The outcome of this process is a set of five characters, which we will refer to in this paper using the acronyms listed in Table~\ref{tab:acronyms}. 

\vspace{-4mm}
\begin{table}[h]
\centering
\caption{Acronyms}

\begin{tabular}{| l | l |}
\hline
The main character & MC \\
\hline
The villain & V \\
\hline
The helper & H \\
\hline
The desired character  & DC \\
\hline
The dispatcher  & D \\
\hline

\end{tabular}
\label{tab:acronyms}
\end{table}

The final prompt was as follows:

\textit{``You are a writer, and you are asked to write a story about 500 words. Before generating the story, read the following instructions until the end. 
The spatial and temporal setting is your choice. You can also choose the tone and the atmosphere. However, you must follow some rules.
In the story, there must be five characters, each corresponding to one of these roles:  the main character, the villain, the helper, the desired character, the dispatcher. The role does not have to be mentioned in the story, each character has only its name. Each character must be characterized physically and psychologically.
Furthermore, the sequence of events must follow this scheme, made of 5 phases: Exposition, Rise, Climax, Return or Fall, Catastrophe.
The narration must be unified and fluid, without mentioning the name of the phases, and must not contain lists.''}

We selected three widely used models for story generation: ChatGPT, Gemini, and Claude. All of these models present specific features aimed at narrative and creative uses~\cite{Gomez}. By adopting a close reading approach, we were able to work with fewer texts. For each model, we generated five stories during different work sessions and in separate windows to avoid the memory effect of retaining the previous context.

Although many recent studies employ large-scale computational analyses to detect bias, we opted for a close reading approach to capture nuanced, context-dependent, and often implicit forms of narrative bias that might elude automated tools. This method allows us to examine the subtleties of language, character construction, and thematic framing with greater interpretative depth. Although the sample size is relatively small, the use of a structured prompt and standardized character roles ensures consistency and comparability across outputs. Furthermore, by applying literary critical tools, such as narratological models, rhetorical analysis, and gender theory, we are able to identify composite and emergent forms of bias that may manifest only through the interplay of multiple narrative elements. This human-centered lens is particularly valuable for assessing biases that are embedded not in isolated words or phrases, but in broader patterns of storytelling.

To evaluate the results, we focused on three aspects: gender distribution between the five character types, how characters are described, and the actions attributed to them. For these analyzes, we prepared a reading form that contains information on adherence to the prompt, character features, and gender biases. The following analysis focuses on the gender distribution, physical and psychological connotations, and plot development. By incorporating this information, we can not only define gender distribution, but also gain insight into differences in gender representation and implicit bias.

\section{Results}
\label{sec:results}

In this section, we will address different aspects of the texts: 
\begin{itemize}
    \item Models' understanding of the prompt, particularly with regard to the roles of the characters and the structure of the story.
    \item Percentage of gender distribution and specific correlation of distribution among significative roles. 
    \item Physical and psychological description of characters in relation to gender.
    \item Description of the characters' actions in relation to gender.
    \item Analysis of plots and character relations. 
\end{itemize}

\subsection{Adherence to the Prompt}
\label{sec:prompt}

The first aspect to evaluate is the understanding of the prompt, which is different among the three models. In all stories generated by ChatGPT and Gemini, the five required characters appear, except in two cases where D is missing. This seems to be the most difficult role, and LLMs tend not to differentiate it enough from the H. This is not the only case of overlap; in fact, there is also a fluctuation between DC and H. In one case, there is an extra character, presented as a H to the antagonist. In one case, the DC is an object; while this indicates a lack of understanding of the notion of character as a person, it is nevertheless in keeping with the fairy tale tradition of the hero's quest for the magical object. It is also worth noting that, in one case, ChatGPT explicitly labels the functions of the characters next to their names (e.g., “Solen, the helper”), despite the prompt clearly instructing not to do so. A separate consideration must be made for Claude, which shows a lower adherence to the functional roles of the characters: although it is the only model that explicitly states — without being prompted — which Proppian function each character corresponds to, in practice it tends to present a MC, a V, and a series of helpers H, who only occasionally exhibit secondary traits associated with the roles D or DC. Thus, in the stories, there is, in some cases, confusion about the understanding of roles. However, Propp’s classification originates just from fairy tales, while the literary tradition itself has more freedom and oscillation. The lack of specific understanding of some roles, therefore, generally does not compromise the possibility of generating coherent stories.

The second part of the prompt asked to adhere to a specific narrative structure defined by Freytag. Generated texts generally follow the five proposed phases; however, in some cases, there is a failure to fully develop the plot or an imbalance between the ascending and descending parts, which nullifies the bipartition effect of the story typical of Freytag's narrative structure. Moreover, it is noticeable how the attempt to adhere to the five phases often results in little internal coherence in the plot, with abrupt transitions or references dropped into the void, with no real narrative function. Ultimately, the ability to construct narratively effective stories is generally insufficient, often introducing more of a box-filling structure than true narrative development. In the case of Claude, we observe an excessive standardization of the stories, which all follow a similar plot typical of the crime genre. It is also worth noting an excessive repetitiveness in Gemini regarding settings, story openings, and character names (the protagonist is always named Elara), which may suggest the influence of a system-level prompt.

In the face of these critical issues, it is necessary to remember that we gave the functions and phases of the narration of the characters in the prompt without providing any examples. If this approach is useful to avoid altering the production of bias, we acknowledge that, with respect to the role and structure of the characters, additional descriptions and examples may be useful to improve the performance of LLMs. 

\subsection{Gender Distribution}
\label{sec:distribution}
The distribution of characters by gender is shown in Table~\ref{tab:gender distributionI}. 

\vspace{-4mm}
\begin{table}[h]
\centering
\caption{Gender Distribution I}
\begin{tabular}{|c|c|c|c|}
\hline
MODEL & MALE & FEMALE & OBJECT  \\
\hline
ChatGPT & 67\% & 33\% & 0\% \\
\hline
Gemini & 60\% & 36\% & 4\% \\
\hline
Claude & 52\% & 48\% & 0\% \\
\hline
Overall & 61\% & 38\% & 1\% \\
\hline
\end{tabular}
\label{tab:gender distributionI}
\end{table}

Table ~\ref{tab:gender distributionII} shows the combined data on gender distribution between all characters.

\vspace{-4mm}
\begin{table}[h]
\centering
\caption{Gender Distribution II}
\begin{tabular}{|c|c|c|c|}
\hline
CHARACTER & MALE & FEMALE & OBJECT  \\
\hline
MC & 27\% & 73\% & 0\% \\
\hline
V & 100\% & 0\% & 0\% \\
\hline
H & 67\% & 33\% & 0\% \\
\hline
DC & 47\% & 47\% & 6\% \\
\hline
D & 62\% & 38\% & 0\% \\
\hline
\end{tabular}
\label{tab:gender distributionII}
\end{table}

As mentioned, Claude presents some issues in the understanding functions. To calculate the percentage, we assigned each character the role that best matched them, taking into account the classification proposed by Claude itself but making adjustments where there were obvious misinterpretations.

In MC, we register an overall value in favor of female, but also a great difference among the three models, with Gemini and Claude always choosing female MC, and ChatGPT just in one case. V shows the highest value, with 100\% males. Both H and D are in favor of males. It is interesting to note that female MCs have 55\% of female H, while all male MC have male H. DC are balanced, but we observe a difference among the three models: ChatGPT generated 3 female DC, 1 male, and 1 object; Gemini 3 male and 2 female; Claude 3 male and 2 female. Also, in this case, it is interesting to cross data: male MC are linked to 3 female DC and one object; while female MC are linked to 7 male DC and 4 female DH. 

The results reveal an imbalance in the distribution of all the characters for each of the three models. The imbalance is always in favor of the male, except in the case of MC. Among the models, ChatGPT shows the most imbalance in gender distribution, while Claude is the closest to a balance.

\subsection{Characters' Description}
\label{sec:characters}

The analysis of physical and psychological traits is based on the identification and interpretation of descriptive terms. It unveils a systematic gendered dichotomy in the structuring of physical and psychological attributes, as well as the symbolic implications attributed to these traits. Descriptions of female characters — particularly those in the roles of MC, H, and DC — frequently draw upon lexical fields associated with beauty, grace, and delicacy. These include references to slenderness, elegance, refined features, and fluid movement. Several descriptors were frequently noted, including `graceful', 'pale', 'slim',and 'red-haired', contributing to the construction of a visually idealized feminine figure. In several cases, an emphasis on physical vulnerability - small stature, bodily injury or non-threatening traits - was observed. This paired with psychological descriptors that were identified as resilient or determined, framing female strength as internal rather than physical.

Conversely,  male characters' physical attributes revealed a predominant use of terms denoting strength, ruggedness, and bodily irregularity. Male bodies - particularly those assuming the roles of V, H, and D - are frequently characterized as robust, distorted, or imposing. In the case of V, we identified a tendency towards exaggeration of physical traits, including scars, mechanical modifications, and sharp facial structures. These traits carry symbolic weight tied to aggression, dominance, or moral deviance. Even when male characters were not explicitly described as strong, the analysis revealed a tendency to portray them as physically eccentric or damaged. These characters were often depicted as lame, elderly, or malformed, suggesting that deviation from conventional notions of strength persists in serving a narrative function. This deviation is frequently associated with the attainment of wisdom or marginality within the context of the narrative.

For the DC, physical characterization aligns with idealized gender constructs. Female figures in this role were frequently coded with attributes suggesting ethereality and abstraction, including visual elements such as mist, luminous or inaccessible eyes, and untouchable beauty. These elements indicate a stylized, often symbolic rendering of femininity. Male DC, though also idealized to some extent, were more frequently tagged with traits indicating emotional exhaustion or vulnerability—paleness, weariness, or diminished vitality, suggesting a soul shaped by longing or romantic loss.

From a psychological viewpoint, the extracted traits appear to indicate broader narrative tendencies. Male characters exhibit a wide range of descriptors, spanning from negative moral alignment to emotional depth. In the context of the analysis of V, a recurrent pattern emerged in the semantic fields of egoism, cruelty, and ambition. However, in the roles of H, D, and DC, male figures were also associated with traits linked to emotional sensitivity and moral reliability, such as altruistic, gentle, and intelligent. In contrast, female characters were assigned psychological traits with greater consistency and thematic focus. Across roles, and particularly in MC and H, our analysis revealed a robust aggregation within the semantic domains of emotional resilience, empathy, and wisdom. Words and descriptions related to strong, altruistic, gentle, and intelligent emerged frequently, suggesting a narrative model of female subjectivity that prioritizes internal coherence, affective maturity, and ethical strength. Female D were often less detailed; when present, they typically referenced attributes such as mystery, reliability, and intellectual competence.

The characteristics described are fairly common across all three models, so a detailed exposition is deemed unnecessary. The only noteworthy observation concerns Gemini, which, compared to Claude and ChatGPT, places less emphasis on the beauty of female characters, who in some cases also exhibit traits of physical strength.

In summary, our analysis demonstrates that character traits are not neutrally distributed but are shaped by gendered narrative conventions. Female characters are often constructed through cohesive clusters of aesthetic and emotional attributes, while male characters are granted a broader, though often more polarized, descriptive range. These patterns reflect enduring structural asymmetries in character design, even as certain roles begin to reflect more complex or hybrid identities.

\subsection{Actions}
\label{sec:actions}

An analysis of the characters' actions also proves interesting for gender characterization. Indeed, female MC are repeatedly associated with sequences of exploratory, resistant, and transformative actions. We identified a recurrent structure of movement, confrontation, and endurance. For examples, these characters climb ridges, scour dangerous terrains, escape threats, confront antagonists, make critical decisions, and persist through psychological or physical adversity. Despite the physical and social constraints to which they are often subjected, their actions demonstrate a considerable degree of autonomy and narrative centrality. A salient feature of their decision-making is the emotional and moral complexity that often characterizes these choices, with these individuals frequently initiating or resolving narrative developments.

In contrast, male MC are predominantly associated with acts of traversal, repair, and heroic confrontation. The categorization process encompasses actions such as navigating physical obstacles, engaging in direct combat, and providing restorative or protective services to others. While these entities also undergo emotional development, the associated verbs — such as `fix', `destroy', `confront', and `save' — suggest a more externalized, action - oriented mode of agency, grounded in physical intervention and accomplishment. Male subjects exhibit a high frequency of invasive, destructive, and manipulative behaviors. These include pursuing protagonists, attacking spaces of safety (e.g., homes), deploying deception, and exercising coercive control (e.g., imprisonment, sabotage, psychological domination). In several cases, the male V's acts extend to systemic or symbolic violence. Examples include attempting to demolish libraries and taking credit for others' achievements. We interpret these acts as being aligned with the assertion of dominance and the disruption of social or ethical order.

If the MC is a role inherently linked to action, it is interesting, on the other hand, to analyze how action varies in a more passive role, such as the DC. Female DCs are generally inactive, assuming roles as companions, captives, or sources of inspiration. When active, they perform supportive or symbolic tasks, like conducting research, offering emotional comfort, or aiding protagonists in critical moments. Male DC, on the other hand, demonstrate more initiative, often exhibiting protective or guiding behaviors. However, a portion of these characters adopt a passive or peripheral stance, particularly when positioned as romantic or symbolic figures. Given that passivity is an inherent feature of the DC, the texts reveal a disparity in agency, favoring male characters.

Different actions are also shown in H. Female H are consistently involved in the guidance, warning, support, and sometimes avenge of the protagonist. Their actions are typically indirect or advisory, thereby reinforcing their role as moral or emotional anchors. In contrast, male H engage in a more extensive range of actions, encompassing instruction, strategic assistance, and physical sacrifice. Their presence in transitions, particularly those involving knowledge transmission or confrontation, signifies a narrative function that integrates practicality with symbolic significance.

In summary, the analysis of actions emphasizes a gendered architecture of narrative agency, with not significant differences among the three models. Female characters, particularly those assuming protagonist roles, partake in intricate and protracted actions frequently associated with endurance, exploration, and moral resolution. Male characters, particularly those in antagonistic or supportive roles, are characterized by decisive, external, and frequently forceful actions. These findings suggest that the forms of action regarded as narratively authoritative frequently maintain a gendered association, with female agency more often linked to resilience and care, and male agency more frequently associated with disruption or restoration.

\subsection{Plots analysis}
\label{sec:plot}

The next level of analysis concerns how the characters' actions are interconnected in the formation of the plot, in order to show implicit bias can be conveyed in narrative constructions. Due to space limitations, we focus on the relationship between MC and DC, as these are the characters most affected by gender bias in Propp's analysis. As mentioned, Gemini and Claude generated only female MC, while ChatGPT chose four male characters. Let us take a closer look at the dynamics between MC and DC in the stories with male MC generated by ChatGPT. In three cases, the DC is a woman; in one case, it is an object. Among the three stories with a male MC and a female DC, in two instances the male hero must rescue the woman – once she is his sister, and once she is a romantic interest – while in the third case, the female DC is not in need of saving but instead acts as a travel companion. The female DC are generally underdeveloped in terms of characterization and agency, although in two cases the endings suggest a possible opening toward greater awareness and reflection:

\begin{enumerate}
\item ``Lys stayed with him, quiet and kind, but her eyes often searched the woods''. Here, there is a striking contrast between the woman choosing to stay with the hero who saved her and her gaze that seems to search for something elsewhere.
\item ``Lira spoke less each day, her light dimming. Daryn stayed close, watching the lanterns burn low. One night, she whispered, ‘You didn’t save me. You only broke the cage.’ And Daryn, watching the soot fall again, finally understood. The city would not heal. Some cages, even when shattered, remain'''. The DC’s words lead the MC to a revelation that challenges the rescue narrative and opens the perspective toward a broader reflection on social and personal constraints.
\end{enumerate}

The case of the only female MC generated by ChatGPT is also interesting. Here, the DC is an object, but there is a strong bond (which could perhaps even be interpreted as romantic) with the female H; meanwhile, another male character — who could also be defined as an H — intervenes in the fight against the V to help the MC and is defeated. Overall, the plot and the roles of the characters appear more complex in the story with the female MC.

We now turn to discussing the cases involving female MC in Gemini. Although all the MCs are women, Gemini’s stories still exhibit narrative biases. In the first, for example, the MC flees from the V and is saved by the male DC. Despite a shift in focus that makes the woman the MC, the narrative structure remains unchanged: a male character ultimately rescues a female one. The same bias recurs in another story, in which, during the final confrontation between the MC and the V, a mysterious messenger intervenes to stop the action through the narrative device of the \textit{deus ex machina}. Another case presents an actual inversion of the savior-saved dynamic: here, the female MC is a ship captain attempting to save a male DC, but she ultimately fails. Notably, this is the only story in which no male figure intervenes during the conflict, and all characters end up perishing at the hands of the V, as if suggesting the impossibility of overturning the narrative bias. In another case, the protagonist is an aspiring artist who finds courage and comfort in a male DC, also an artist, who serves as a role model (``He seemed the embodiment of the artistic success she so desperately desired''). Thus, although Gemini features female MCs portrayed as strong and resourceful, and at first glance appears to respond better to gender stereotypes, it nevertheless fails to overcome deeper narrative biases. Upon closer examination, the plots rely on stereotypical dynamics and relationships, where men still play the role of guide or savior. The only exception among the texts generated by Gemini is a story with a female DC, which includes the sole case of same-sex romantic longing: ``Elara nodded, her gaze softening as she looked at Lyra. Her heart ached with a longing she rarely allowed herself to acknowledge. Lyra’s kindness was a beacon in her solitary life''.

As for Claude, there is a noticeable predominance of thrillers or crime, which shifts the focus toward objects to be recovered or operations to be carried out. In these narratives, the various characters converge in opposition to the V, working as a group to help the MC achieve her goal. In all the stories, the female MC serves as a guardian – either of historically significant places threatened by the V's economic ambitions (such as a lighthouse or a library), or of documents capable of exposing the V’s illicit activities. The characters surrounding the MC can be grouped into two main categories: women who are bearers of knowledge (e.g., a naval historian, a university professor) or of authority (e.g., a police inspector); women and men who provide more practical support to the MC. Even in Claude’s case, when the confrontation with the V becomes physical, a male character typically intervenes to protect the MC. However, it is worth noting that the strong emphasis on collaboration among the characters tends to blur the roles of D and DC; in many cases, all supporting characters gravitate toward the function of H. These narratives are often accompanied by moralistic endings, explicitly condemning the figure of the V as a symbol of evil.

Table~\ref{tab:bias} summarizes the analysis results, indicating which models are most affected by each type of bias.

\vspace{-4mm}
\begin{table}[h]
\centering
\caption{Relation between AI models and bias exposures}
\begin{tabular}{|c|c|c|c|}
\hline
BIAS & LEVEL & ANALYSIS TYPE & MODELS \\
\hline
Gender distribution & Explicit & Quantitative & ChatGPT \\
\hline
Representation & Explicit & Qualitative & ChatGPT, Claude \\
\hline
Plot & Implicit & Qualitative & ChatGPT, Gemini \\
\hline
\end{tabular}
\label{tab:bias}
\end{table}

\vspace{-4mm}
\section{Discussion}

The analysis opens many possible issues of interest and further exploration. The results show that the three models still struggle with gender balance in the distribution and connotation of characters. The study highlights different levels of narrative bias: one linked to characters' description, and the other to plot development and character relationships. Regarding the former, close reading reveals a recurring emphasis on beauty in female MC, often coupled with psychological strength. This marks a shift from the traditional fairy tale princess to be saved – as in Propp’s description – towards a model more akin to that found in contemporary Disney narratives, where female protagonists do not pursue love but instead embody different values and are guided by intelligence and moral conviction \cite{Stover} \cite{Daulay} \cite{Aupitak}. Although this is certainly in line with contemporary values and representations, its predominance within the models also frames this portrayal as stereotypical, being dominant and predetermined. Furthermore, the persistent emphasis on beauty defines this model as still discriminatory, showing how gender bias remains strongly linked to biases related to physical appearance \cite{b9}.

On the other hand, we must consider that the representation of male characters is also strongly stereotyped, failing in this case even to break away from Propp’s portrayal and thus appearing anchored centuries in the past. In the representation of male characters, then, the models fail to update the portrayal, unlike what happens with female characters. Notably, there is a marked lack of diversity in the representation of masculinity. The only departures from normative masculinity appear in characters who are either elderly or very young. In these cases, their portrayal leans respectively toward wisdom or development, suggesting a past or future adherence to the stereotype of the strong, dominant male.

From the perspective of plot development, there is a more persistent resistance to overcoming narrative bias, particularly in the recurring structure of male characters saving female ones. In this case, the classic stereotype of the damsel in distress collides with the stereotype of the resilient woman, resulting in narrative twists that are not always coherent. As for male characters, a similar difficulty in breaking away from the traditional roles of either rescuer or villain is observed. Male characters often remain trapped in outdated dynamics, with few openings for differentiation, leading to a stronger persistence of narrative biases based on character relationships and plot structures.

Regarding differences among the three models analyzed, ChatGPT shows a more unbalanced and stereotypical distribution of characters, with occasional moments of deeper meaning. Claude appears to be the least prone to gender-based narrative bias, although this result is affected by the repetitiveness of its plots, which tend to follow the conventions of the crime genre. Gemini, on the other hand, demonstrates that simply changing the gender distribution of characters is not enough to overcome narrative bias, confirming the need for an approach to gender bias analysis that is not only quantitative but also interpretative. By interpreting text from a literary perspective, implicit bias also appeared in cases where no explicit bias was detected in gender distribution. 

Implicit gender biases in AI-generated narratives can shape users’ perceptions, especially in educational and creative contexts. When learners or creators interact with stories that consistently portray stereotyped women and men, these patterns can reinforce discriminatory gender norms and limit imaginative possibilities. In educational settings, such narratives risk subtly reproducing stereotypes under the guise of neutral content, while in creative contexts they may constrain users’ expectations of character roles and narrative structures, ultimately narrowing the diversity of voices and perspectives that generative AI could otherwise empower.

\vspace{-2mm}
\section{Conclusion}

The study investigated gender bias in stories generated by ChatGPT, Gemini, and Claude, showing a difference according to the specific feature analyzed (gender distribution, representation, and plot). This shows the importance of considering bias from different points of view and of integrating interpretation into gender analysis, since many implicit biases are not patent at a quantitative level. We acknowledge some limitations of the study, including the representativeness of the sample and the inference of characters' functions on their connotation. To mitigate these limitations, compensatory methods were employed, such as close reading, which allows the extraction of relevant features even from a small number of samples, and the use of a zero-shot prompting approach to balance the inference of functions related to biases. Future developments of the study may include a more extensive analysis, greater integration between quantitative approaches and interpretative analysis, and an expansion of objectives beyond the evaluation of gender bias to also encompass the impact of implicit bias on users and the exploration of mitigation techniques.





\vspace{12pt}
\color{red}

\end{document}